\documentclass[sigconf]{acmart}

\usepackage{booktabs} 
\usepackage{multirow}
\usepackage[belowskip=-15pt,aboveskip=0pt]{caption}
 
\setcopyright{rightsretained}
\begin{document}





\copyrightyear{2019} 
\acmYear{2019} 
\setcopyright{acmcopyright}
\acmConference[EASE '19]{Evaluation and Assessment in Software Engineering}{April 15--17, 2019}{Copenhagen, Denmark}
\acmBooktitle{Evaluation and Assessment in Software Engineering (EASE '19), April 15--17, 2019, Copenhagen, Denmark}
\acmPrice{15.00}
\acmDOI{10.1145/3319008.3319354}
\acmISBN{978-1-4503-7145-2/19/04}

\title{Modern code reviews - Preliminary results of a systematic mapping study}

\author{Deepika Badampudi}
\affiliation{%
  \institution{Blekinge Institute of Technology}
  \department{Software Engineering Research Lab}
  \city{Karlskrona}
  \country{Sweden}
}
\email{deepika.badampudi@bth.se}

\author{Ricardo Britto}
\affiliation{%
  \institution{Ericsson AB}
  \institution{Blekinge Institute of Technology}
  \department{Software Engineering Research Lab}
  \city{Karlskrona}
  \country{Sweden}
}
\email{ricardo.britto@[ericsson.com,bth.se]}

\author{Michael Unterkalmsteiner}
\affiliation{%
  \institution{Blekinge Institute of Technology}
  \department{Software Engineering Research Lab}
  \city{Karlskrona}
  \country{Sweden}
}
\email{michael.unterkalmsteiner@bth.se}


\begin{abstract}
Reviewing source code is a common practice in a modern and collaborative coding environment. In the past few years, the research on modern code reviews has gained interest among practitioners and researchers. The objective of our investigation is to observe the evolution of research related to modern code reviews, identify research gaps and serve as a basis for future research. We use a systematic mapping approach to identify and classify 177 research papers. As preliminary result of our investigation, we present in this paper a classification scheme of the main contributions of modern code review research between 2005 and 2018. 
\end{abstract}

%
%



\maketitle
\section{Introduction}
Code reviews have evolved from being rigorous and intensive to lightweight and collaborative. Modern code reviews are conducted to examine the changes made to a software system and to evaluate its quality. In an open source or inner source project, people other than the core team can make contributions to a software product. Hence, reviewing the code before it is accepted and merged is crucial; not necessarily to identify faults, but rather to improve solutions, share knowledge and code ownership~\cite{Bacchelli2013}. 

Although modern code review has become a prevalent practice in the software industry, there is no study that aggregates existing associated literature and identifies gaps in the body of knowledge. Therefore, the aim of our investigation is to conduct a mapping study to aggregate existing literature in this area and identifying gaps. In this paper, we present the preliminary results of our investigation. We followed the guidelines by Petersen et al.~\cite{PETERSEN20151} to conduct a systematic mapping study. The contributions of our mapping study are to provide a classification scheme of modern code review research, evolution of research topics and identify research gaps and potential future research areas. 

The structure of the paper is as follows: Related work is described in Section~\ref{sec:related work}. The research design used to conduct systematic mapping study is described in Section~\ref{sec:SMS process}. The preliminary results are presented in Section~\ref{Sec:results}. Finally, our conclusions and next steps in our investigation are presented in Section~\ref{Sec:future work}.

\section{Related work} \label{sec:related work}
Nargis et. al~\cite{Nargis2018} have published a systematic literature review protocol, designed at identifying challenges and benefits of modern code reviews. Our study has a broader scope, i.e. we intend to map all research on modern code reviews without limitation on outcome. Earlier reviews target traditional code inspections~\cite{Tenorio2016} or peer assessment outside the software engineering domain~\cite{hernandes2013overview}. To the best of our knowledge, no systematic investigation on what we know about modern code reviews has yet been conducted.

\section{Research design} \label{sec:SMS process}
The guidelines by Petersen et al.~\cite{PETERSEN20151} include the following steps: (1) definition of review questions; (2) conduct search for primary papers; (3) screening relevant papers; (4) keywording of abstracts; (5) Data extraction and mapping of studies. 
\subsection{Goal and review questions}
\textbf{Goal}: The main goal of this review is to provide an overview of the existing research on modern code reviews. The overview consists of the contributions of the different research articles and the research approach used in the papers. The purpose of the review is to map the frequencies of research published in the area of modern code reviews to observe the evolution of the research topic. 
Based on the goal of the review, we have formulated the following review questions:
\begin{enumerate}
    \item [R1] What aspects/topics of modern code reviews are investigated?
    \begin{enumerate}
        \item[R1.1] How have the aspects/topic changed over time?
        \item[R1.2] How many articles cover the different aspects of modern code reviews?
    \end{enumerate}
    \item [R2] How were the aspects in R1 investigated? 
\end{enumerate}

Note that the preliminary results presented in this paper only address RQ1.

\subsection{Search strategy}
We employed the following search strategy: \\
\textbf{Databases included}: After defining the review questions, the next step is to select databases to find the relevant papers. The following databases were selected based on their coverage of papers: Scopus, IEEE Explore, and ACM Digital Library.\\
\textbf{Search string}: In order to search for relevant papers in the three databases, we used the following keywords: {Code review, Modern code review, Contemporary code review, Patch accept, Commit review, Pull request, Modern code inspect}. We used the keywords in the search engines using the "OR" operator between each keyword and the keywords were suffixed with a wildcard - "*".
   The result of applying the search strings on the three databases is presented in Table~\ref{tab:database resutls}.
    \begin{table}
        \caption{Search results from each database}
        \label{tab:database resutls}
        \begin{center}
        \begin{tabular}{ll}
        \toprule
           \textbf{Database}  & \textbf{Papers} \\
           \midrule
            Scopus & 866 \\
            IEEE Explore & 335\\
            ACM Digital Library - Title & 99\\
            ACM Digital Library - Abstract & 243\\
            \midrule
            \textbf{Total} & 1543 \\ 
            \textbf{Total after removing duplicates} & 873 \\
            \bottomrule
        \end{tabular}
        \end{center}
    \end{table}
    
   \subsection{Inclusion-exclusion criteria} \label{Sec:incl/excl}
     The 873 identified papers were reviewed based on a defined set of inclusion and exclusion criteria. Before we started the review process, we conducted a pilot study\footnote{In our first pilot study, we noticed a paper on test case review, we refined our inclusion criteria 1 to add test code review as well. Some papers discussed approaches to support the modern code review process to make it more efficient, for example, by selecting a relevant reviewer. Therefore, we added inclusion criteria 2. We decided to exclude papers that discuss modern code reviews in education. We modified exclusion criteria 1 to emphasize the subject of the investigation; we only include papers where the process of modern code review is under investigation. We also came across papers that discuss solutions that might benefit, among other things, the modern code reviews process, without discussing the implications of the approach on the code review process itself (e.g., defect prediction). As a result, we excluded such papers and added exclusion criteria 2.} on 20 papers to ensure that all the authors have the same interpretation of the criteria.
   After the pilot study, the initial criteria were updated and new criteria were added. We conducted a second pilot study on 20 additional papers using the revised criteria. As a result, we achieved higher consensus in our decision. The final formulation of the criteria is as follows:
    \paragraph{Inclusion criteria}
    \begin{enumerate}
        \item Papers discussing source code (including test code) review which is done on regular basis (modern code review done on every pull request to either accept/reject them).
        \item Papers discussing aspects such as reviewer selection, what code to review, etc. that support modern code review process.
        \item Papers discussing reviewer and/or developer perspective.
        \item Papers (peer reviewed and grey literature) related to modern code reviews.
        \item Papers including modern code reviews from different aspects. Examples - Benefits, outcomes, challenges, motivations, quality, usefulness and so on. 
        \item Papers proposing solutions for modern code review.
    \end{enumerate}  
    \paragraph{Exclusion criteria}
    \begin{enumerate}
        \item Papers not discussing modern code reviews or the subject of investigation is not the modern code review process.
        \item Papers that do not discuss the implications of a solution on modern code review process.
        \item Papers that discuss modern code review in education.
        \item Papers not in English and those that do not have full text available.
     \end{enumerate}


\subsection{Keywording of abstracts and meta-data}
During the screening process, we looked at the abstracts to find keywords and concepts that represent the contribution of the papers. 
We collected the following data from the selected papers:
\begin{itemize}
    \item \textit{Overview of the main contribution} - The contribution could be related to the different aspects of modern code review, solutions for modern code review improvement or discussion of modern code review process from reviewer/developer perspective.
    \item \textit{Author} - The authors of the papers.
    \item \textit{Publication type} - Conference, journal or book.
    \item \textit{Year} - Year of publication.
\end{itemize}

\subsection{Data extraction}
In addition to the data extracted during the keywording process, we will extract research facet based on Wieringa's \cite{Wieringa:2005:REP:1107677.1107683} classification. We will extract the information needed to evaluate the rigor and relevance of the papers. Moreover, the data extracted through the keywording process will be refined (if necessary) based on full-text reading.  

\subsection{Validity threats}
It is important to address the validity threats relevant to a mapping study which are as follows: \\
\textit{Researcher bias in inclusion/exclusion} - All three authors were included in the screening process. We conducted two pilot studies to ensure that all authors had the same interpretation of the inclusion/ exclusion criteria. In case of doubts, we discussed the papers together and revised the criteria to make them more explicit (see Section~\ref{Sec:incl/excl}).\\
\textit{Exclusion of relevant papers} - We adopted an inclusive approach; whenever we were in doubt regarding a paper, we included it for further reading. We marked all excluded papers with the applicable exclusion criteria to ensure transparency and traceability.

\section{Review results} \label{Sec:results}
In this section, we present our preliminary results, i.e. answers to R1. In total, we have included 177 papers after screening the abstracts\footnote{We did not include the 21 papers selected in the pilot study and 46 tentatively accepted papers in this paper due to the space constraints. We will include them in our extended paper}.

\begin{figure}
    \centering
    \includegraphics[width=\columnwidth]{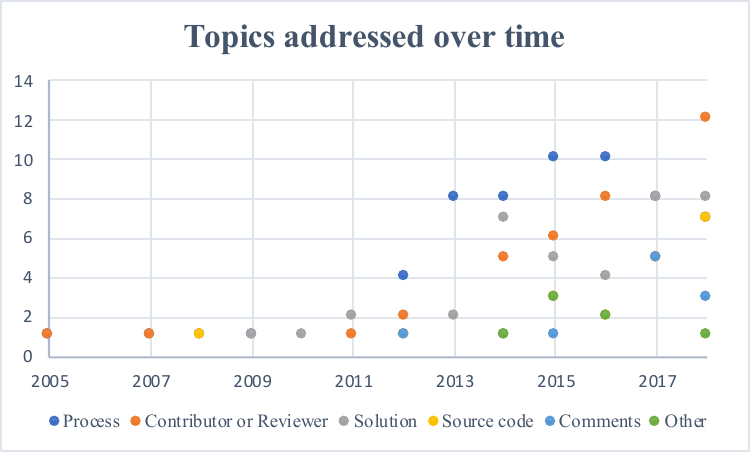}
    \setlength{\belowcaptionskip}{10pt}
    \caption{Number of papers published per year}
    
    \label{fig:year}
\end{figure}

Table~\ref{tab:Aspects} presents the aspects (R1) and the number of papers that cover the different aspects of modern code reviews (R1.2). The most investigated topics are "solutions", "impact/outcome", and "reviewer identification". We elaborate on these topics in Sections~\ref{Sec:solution}, \ref{Sec:impact} and~\ref{Sec:revid} respectively. To investigate how the research topics evolved over time (R1.1), we first split all the papers by the year in which they were published. Figure~\ref{fig:year} shows the number of papers published per year (2005-2018). In the last five years, the number of papers has increased drastically compared to previous years.

\begin{table}
\caption{Studied aspects of modern code reviews}\label{tab:Aspects}
\begin{tabular}{p{1.3cm}p{5.5cm}p{0.3cm}}
\toprule
Aspects  & Sub-Aspects & No.\\ 
\midrule
\multirow{6}{*} {\begin{tabular}[c]{@{}l@{}}Modern \\ code \\ review \\ process\end{tabular}}  & Benefits of modern code review  & 3 \\
    & Causes - acceptance/ rejection/ partial acceptance/ integration delays of pull requests & 6 \\
    & Motivations, expectations, challenges and/ or best practices & 4 \\
    & Characteristics / principles of modern code review processes & 8\\
    & Effectiveness and/or efficiency of modern code review & 3\\
    & Impact/outcome & 35\\
    \midrule
\multirow{3}{*}{\begin{tabular}[c]{@{}l@{}}Contributor\\/Reviewer\end{tabular}}   & Perception on modern code review & 3\\
    & Characteristics such as skills, behaviour and/or participation & 15\\
    & Reviewer selection & 23\\
    \midrule
    Solution & Tool support & 39\\ 
    \midrule
\multirow{2}{*}{\begin{tabular}[c]{@{}l@{}}Source \\ code\end{tabular}}           & Code characteristics & 5\\
    & Identification of code to review & 15\\
    \midrule
\multirow{3}{*}{\begin{tabular}[c]{@{}l@{}}Review \\ comments\end{tabular}}       & Classification of comments & 3\\
    & Assessment of comments & 3\\
    & Usefulness of the comments & 5\\
    \midrule
    Other & &   7\\
    \bottomrule
\end{tabular}
\end{table}

\subsection{Papers proposing solution to improve modern code review process} \label{Sec:solution}
We have identified 39 solution papers in our mapping study. A few papers proposed solution for the same purpose: support for collaborative modern code review~\cite{raab2011collaborative,shochat2008seecode,muller2012approach,henley2018cfar}, identifying behavior-modifying changes~\cite{alves2014refdistiller,ge2014towards}, motivation enhancement~\cite{unkelos2016lets,unkelos2015gamifying}, use of  static analysis to reduce modern code review effort~\cite{singh2017evaluating,balachandran2013reducing}, and the information needs of modern code review~\cite{pascarella2018information}. Table~\ref{tab:Solutions} provides a list of solutions addressing different purposes. 

\begin{table}
    \caption{Proposed solutions}\label{tab:Solutions}
    \begin{tabular}{p{7.5cm}l}
    \toprule
    Purpose & Ref.\\
    \midrule
Improve the skills of newcomer developers    & \cite{oeda2018development} \\
Measure the performance of modern code review and a method to study linux-style reviews    & \cite{izquierdo2017using} \\
Address challenges in modern code review process when conducted by third-part organization    & \cite{ishida2018visualization} \\
Allow developers to explain their code using voice and video    & \cite{hao2013mct} \\
Remove stagnation from modern code review    & \cite{viviani2016removing} \\
Analyse how code changes affect test code    & \cite{oosterwaal2016visualizing} \\
Assist mining modern code review data by enabling better understand of dataset context and identifying abnormalities    & \cite{thongtanunam2014reda} \\
Automatically identify factors that confuse reviewers    & \cite{ebert2017confusion} \\
Automatically partition composite changes and then facilitate modern code reviews    & \cite{tao2015partitioning} \\
Cleaner modern code review    & \cite{lal2017code} \\
Modern code review driven by software quality concerns    & \cite{tymchuk2015treating} \\
Collect modern code review data, generate metrics and provide ways to access the metrics and data    & \cite{bird2015lessons} \\
Community-based modern code review    & \cite{zhang2011design}  \\
Determine how code changes should be ordered to facilitate modern code reviews (increase cognitive support)    & \cite{baum2017optimal}  \\
Estimate  modern code review effort based on patch size and complexity    & \cite{mishra2014mining} \\
Explore the need for a new generation tool of modern code review tools    & \cite{baum2016need} \\
Focus on design quality concerns    & \cite{tymchuk2015vidi} \\
Choose pre- or post-commit modern code reviews    & \cite{baum2017comparing} \\
Identify problems with modern code review processes    & \cite{10.1007/978-3-642-55128-4_1} \\
Identify security problems with web-based systems    & \cite{cap2018ensuring} \\
Improve modern code review process within organization    & \cite{czerwonka2018codeflow} \\
Peer modern code review    & \cite{holzmann2010scrub} \\
Retrieve modern code review data    & \cite{gonzalez2014analyzing} \\
Summarize similar changes and detect missing or inconsistent edits    & \cite{zhang2014critics} \\
Support the modern code review of visual programming languages    & \cite{ragusa2018code} \\
Track modern code review performance & \cite{izquierdo2018software} \\
Use of mobile to review source code    & \cite{frkacz2017experimental} \\
Use of social networks for frequent modern code reviews    & \cite{durschmid2017continuous} \\ 
\bottomrule
    \end{tabular}
\end{table}

\subsection{Impact and/or outcome} \label{Sec:impact}
The impact of modern code reviews is one of the frequently investigated topics. The topics investigated are impact on software quality~\cite{5, 8, 9, 12, 14}, human memory~\cite{3, 4}, chances in inducing bug fixes~\cite{5}, if-statements and change requests~\cite{6, 7, 10}, and the identification of bugs/warning/vulnerabilities~\cite{11, 13}. 

While the impact of modern code reviews on several factors was investigated as mentioned above, the impact of several factors on modern code reviews has also been investigated: the impact of continuous integration~\cite{17}, developer reputation~\cite{18}, geographical location~\cite{19}, pair programming~\cite{20}, patch voting~\cite{21}, and technical and non-technical factors~\cite{29} on modern code reviews. 

The relationship between specific characteristics of modern code reviews and their respective impact on different aspects of software development has also been investigated (Table~\ref{tab:impact}).  

\begin{table}
    \caption{Relationship between different factors investigated}
    \label{tab:impact}
    \begin{tabular}{p{3.5cm}p{3.5cm}p{0.5cm}}
    \toprule
    Impact of & Impact on & Ref. \\ 
    \midrule
       Code ownership and reviewer expertise  & Software quality & \cite{2} \\
        Reviewer age and efficiency  & Software quality & \cite{28} \\
       Reviewers disagreement & Software quality & \cite{23} \\
       Modern code reviews feedback & Motivation to contribute in OSS projects & \cite{15}  \\
       Continuous code review process & Understandability and collective ownership of the code base & \cite{16} \\
       Reviewers' reviews and personal and social factors & Review quality & \cite{27} \\
       Peer-Based Software Reviews & Team Performance & \cite{22} \\
       Socio-Technical modern code review metrics & Identification of bugs/warning/vulnerabilities & \cite{24} \\
       Reviewers' path development experience & Identification of bugs/warning/vulnerabilities & \cite{26} \\
       Feedback & Developers' sentiment & \cite{25} \\ \bottomrule
    \end{tabular}
\end{table}

The impact of non-technical factors (different patch size, patch priority, component, reviewer load, reviewer activity, and patch writer experience) was investigated in~\cite{31}. The outcome of modern code reviews in terms on detection of code smells~\cite{32} and defects~\cite{33}, ability to identify information on design decisions~\cite{34} and design rational~\cite{36} and ability to discover misunderstandings about object oriented principles~\cite{35} have also been investigated. 
    
\subsection{Reviewer identification} \label{Sec:revid}
Papers proposing solutions to recommend modern code reviewers based on different selection criteria are presented in Table~\ref{tab:reviewersel}. 
    \begin{table}

        \caption{Reviewer selection criteria}
        \label{tab:reviewersel}
        \begin{tabular}{p{5.5cm}p{2cm}}
        \toprule
        Reviewer selection criteria & Ref. \\ \midrule
        Previously reviewed file path &\cite{thongtanunam2015should, thongtanunam2014improving} \\
        Reviewers expertise/experience  & \cite{ouni2016search, hannebauer2016automatically, balachandran2013reducing, rahman2016correct}\\
        Reviewer comment networks &\cite{yu2014should}\\
        Modern code reviewer profiles &\cite{fejzer2018profile}\\
        Participation in the core team & \cite{de2015developers}\\
        Reviewers' activeness  &\cite{jiang2017should, yang2017empirical} \\
        Social network analysis &\cite{yu2014reviewer}\\
        Text and file location &\cite{xia2015should}\\
       Topic modelling of historic source code changes and reviews & \cite{kim2018understanding} \\
        File change history, relationships between contributors and activity & \cite{jiang2015coredevrec} \\
        Developer expertise, text similarity and social relations & \cite{ying2016earec} \\
        Code contributions & \cite{zanjani2016automatically} \\
        Collaborators & \cite{ouni2016search, liao2017topic} \\ \bottomrule
        \end{tabular}
    \end{table}

A solution recommending the role (managerial or technical) of the review was proposed in~\cite{yang2018revrec}. The motivation of invited reviewers was investigated and led to proposing guidelines for inviting reviewers~\cite{ruangwan2018impact}. The factors that influence reviewer assignment was investigated in~\cite{SOARES201832}. The use of reviewer recommendation was investigated in~\cite{kovalenko2018does} and~\cite{peng2018exploring}. 

The results indicate that, even though reviewer selection is perceived to be relevant and effort saving, it rarely adds additional value and creates an unbalanced work load. 

\section{Conclusions and next steps} \label{Sec:future work}
In this paper we presented the preliminary results of a systematic mapping study on code reviews. We have included 177 papers and extracted data from their abstracts to answer our RQ1 research question. We identified a steady upward trend regarding publications related to modern code review starting in 2011. Moreover, the main aspects addressed by existing research are related to: modern code review process, reviewer selection, tool support, identification of code to be reviewed, and analysis of the review comments.

The following are the next steps in our investigation: i) describe all the aspects of modern code review in Table \ref{tab:Aspects} in detail (e.g., the techniques used to propose solutions). This will give an overview on how the solutions have been designed and what are the limitations, gaps, and potential future works; ii) measure the rigor and relevance of the included studies. Such analysis can determine the strength of the results; iii) extract additional data from the papers' full text (e.g., research facet) and verify the classification of the aspects that was extracted from the abstracts; iv) include a detailed discussion of the results.

\bibliographystyle{abbrvnat}
\bibliography{References}

\end{document}